\documentclass[twocolumn,pra,superscriptaddress]{revtex4}
\makeatletter

\newcommand{\Rmnum}[1]{\expandafter\@slowromancap\romannumeral #1@}
\makeatother
\usepackage{amsmath}
\usepackage{graphicx}
\usepackage{subfigure}
\usepackage{color}
\usepackage{amsfonts}
\usepackage{tikz}
\usepackage{esint}
\usepackage{mathrsfs}
\usepackage{hyperref}
\hypersetup{
     colorlinks = true,
     citecolor  = blue,  
     linkcolor  = red 
}
\begin{document}
\title{Emulating topological currents arising from a dipolar parity anomaly in two-dimensional optical lattices}

\author{Zhi Lin}
\affiliation{School of Physics and Materials Science, Anhui University, Hefei 230601, China}
\affiliation{Department of Physics and Center of Theoretical and Computational Physics, The University of Hong Kong, Pokfulam Road, Hong Kong, China}
\affiliation{Shenzhen Institute of Research and Innovation, The University of Hong Kong, Shenzhen 518063, China}

\author{Xian-Jia Huang}
\affiliation{Department of Physics and Center of Theoretical and Computational Physics, The University of Hong Kong, Pokfulam Road, Hong Kong, China}

\author{Dan-Wei Zhang}
\email{danweizhang@m.scnu.edu.cn}
\affiliation{Guangdong Provincial Key Laboratory of Quantum Engineering and Quantum Materials,
SPTE, South China Normal University, Guangzhou 510006, China}

\author{Shi-Liang Zhu}
\affiliation{National Laboratory of Solid
State Microstructures and School of Physics, Nanjing University,
Nanjing 210093, China}
\affiliation{Guangdong Provincial Key Laboratory of Quantum
Engineering and Quantum Materials, SPTE, South China Normal
University, Guangzhou 510006, China}

\author{Z. D. Wang}
\email{zwang@hku.hk}
\affiliation{Department of Physics and Center of Theoretical and Computational Physics, The University of Hong Kong, Pokfulam Road, Hong Kong, China}
\affiliation{Shenzhen Institute of Research and Innovation, The University of Hong Kong, Shenzhen 518063, China}
\affiliation{Guangdong Provincial Key Laboratory of Quantum
Engineering and Quantum Materials, SPTE, South China Normal
University, Guangzhou 510006, China}

\begin{abstract}
Dipolar parity anomaly can be induced by spatiotemporally weak-dependent energy-momentum separation of paired Dirac points in two-dimensional Dirac 
semimetals. Here we reveal topological currents arising from this kind of anomaly. A corresponding lattice model is proposed to emulate the topological currents by using two-component ultracold atoms in a two-dimensional optical Raman lattice. In our scheme, the topological currents can be generated by varying on-site coupling between the two atomic components in time and tuned via the laser fields. Moreover, we show that the topological particle currents can directly be detected from measuring the drift of the center of mass of the atomic gases.
\end{abstract}

\maketitle

\section{Introduction}

Topological states of quantum matter have been paid significant attention in condensed-matter physics \cite{Kane,Zhang_SC} and belong to a new classification paradigm based on the notion of topological order \cite{Thouless,Wen_XG}, which is distinctly different from the states that are characterized by Landau theory based on spontaneous symmetry breaking \cite{SSB}. In recent years, more and more interest have shifted from gapped topological insulators or superconductors to gapless topological systems, including topological semimetals with Dirac points \cite{D_point1,D_point2,D_point3}, Weyl points \cite{W_point1,W_point2,W_point3,TaAs1,TaAs2,NbAs,Ca3P2,PbTaSe21,PbTaSe22,ZrSiS},  Dirac line nodes \cite{D_line1,D_line2,D_line3,D_line4,D_line5,carbon1,carbon2}, as well as symmetry-protected $\mathbb{Z}_{2}$-type gapless points \cite{zhao2013,zhao1} and spin-1 Maxwell points \cite{ZYQ2017,Tan2018}.

On one hand, the topological electromagnetic responses have  emerged as a hot topic of research on topological semimetals. For three-dimensional (3D) topological Weyl semimetals, there exists a topological current $\mathbf{J}_{W}=b_{0}\mathbf{B}/4\pi^{2}$ arising from the chiral magnetic effect in the presence of the energy separation of paired Weyl points $b_{0}$ and the external magnetic field $\mathbf{B}$ \cite{CME}, which has been virtually simulated with superconducting quantum circuits in 3D parameter space \cite{SQC2}. Another kind of topological responses is induced by the so-called axial gauge fields coupling with opposite signs to Weyl fermions of opposite chirality, which emerge due to spatially and temporally dependent separations of a pair of Weyl points in energy-momentum space \cite{CXLiu,Chernodub,Grushin,ZMHuang,axial_GF}. In two-dimensional (2D) graphene, similar gauge fields coupling with opposite signs to each Dirac point (valley) can emerge from an external strain field \cite{Amorim}. Unlike Weyl points in Weyl semimetals, the topologically stable Dirac points in 2D systems should be protected by certain robust kinds of  symmetry. Therefore, topological responses (currents) in 2D Dirac semimetals related to quantum anomalies remain unclear. Notably, a topological edge current may be obtained from time-dependent elastic deformations in gapped graphene preserving time-reversal symmetry \cite{Vaezi}, which lies on a mixed Chern-Simons term in the effective action that involves the electromagnetic and the elastic vector potentials.

On the other hand, ultracold atoms in optical lattices provide a powerful platform to simulate various quantum states of matter due to their high flexibility and controllability \cite{QSimu1,QSimu2,QSimu3,AGF,ZDW2018}. Remarkably, various topological systems have been realized and measured with ultracold atoms, such as the Hofstadter model \cite{hofstadter,Measuring-chern-number}, Haldane model \cite{haldane-model}, 2D spin-orbit-coupled systems with Dirac points \cite{2D-SOC-Dirac} and with nonzero Chern numbers \cite{Liu2014,2D-SOC}, and the chiral edge states \cite{chiral-edge-states}. Moreover, several schemes have been proposed to realize various topological semimetal bands \cite{Jiang,Dubcek,ZDW2015,He,Xu1,ZDW2016,HPHu} 
and artificial (axial) gauge fields \cite{axial_GF,ZDW2014,BBTian} with cold atoms in optical lattices. However, feasible cold-atom schemes for the experimental realization (simulation) and detection of topological currents  are still badly needed.

In this paper, we reveal a distinct kind of topological currents of dipolar parity anomaly in 2D $\mathbb{Z}_{2}$-type semimetals possessing a joint space-time inversion ($PT$) symmetry~\cite{zhao1}. 
Such topological currents in the bulk are induced by the spatiotemporally dependent energy-momentum separation of paired Dirac points, which leads to an effective gauge field that can be viewed as a 2D analog of axial gauge fields.
We also propose a tunable lattice model, which is theoretically able to generate pure topological currents and is realizable by using two-component ultracold atoms in a 2D optical Raman lattice. In the proposed optical lattice system, the topological currents can experimentally be generated by varying on-site coupling between the two atomic components in time and tuned via the applied laser fields. Furthermore, we show that the topological particle currents can directly be detected from measuring the drift of the center of mass of the atomic gas.

The rest of the paper is organized as follows. In Sec. II, we present the general results of topological currents in 2D Dirac semimetals. Section III introduces the tight-binding model for realizing tunable topological currents with cold atoms in a 2D optical Raman lattice. In Sec. IV, we propose schemes for detecting the topological particle currents in the optical lattice. Finally, a brief discussion and a short summary are given in Sec. V.

\section{Topological currents in 2D Dirac semimetals}

\begin{figure}[h!]
\includegraphics[width=0.9\linewidth]{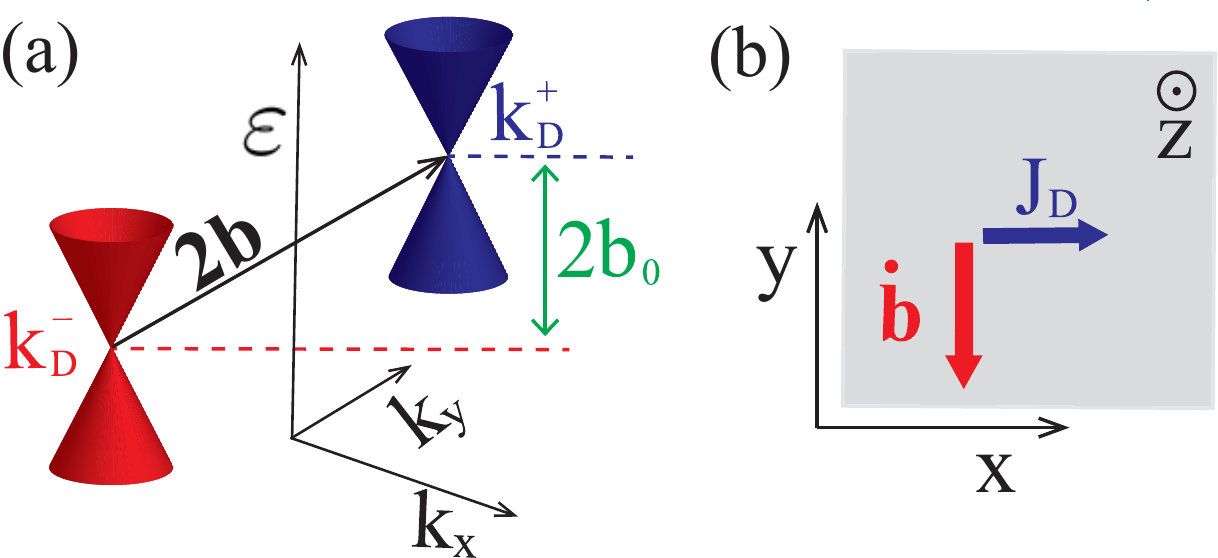}
\caption{(a) A sketch of the dipole momentum $b_\mu=(b_0,\mathbf{b})$ of paired Dirac points in energy-momentum space for $b_{1}=0$, where $b_{\mu}$ is a function of space and time. The Dirac points $\mathbf{k}_{D}^{+}$ and $\mathbf{k}_{D}^{-}$ are, respectively, in blue and red Dirac cones, with $\pm$ denoting the two opposite parity signs. (b) A sketch of the topological current induced by the dipolar parity anomaly $\mathbf{J}_{D}$ from a weak time-dependent $\mathbf{b}$ perturbation.}
\label{Dirac_cone}
\end{figure}

We consider a 2D Dirac semimetal system with a single pair of Dirac points $\mathbf{k}_{D}^{\pm}$, as shown in Fig. \ref{Dirac_cone}(a). The low-energy 
effective Hamiltonian of the system reads 
\begin{equation}\label{HamEff}
\mathcal{H}_{\text{eff}}=k_{\mu}\gamma^{\mu}-b_{\mu}\gamma^{\mu}\tau_{z}
\end{equation}
with $\mu=0,1,2$, where $\gamma^{\mu}$
are the Gamma matrices defined here in the (2+1) dimension of space-time as $\gamma^0=\sigma_0  \tau_0$,  $\gamma^1=\sigma_1 \tau_0$, and $\gamma^2=\sigma_2 \tau_z$, with $\sigma_{\mu}$ being the Pauli matrices
acting on the (pseudo-)spin states at each Dirac cone, while $\tau_0$ and $\tau_z$ belong to the other set of Pauli matrices acting on the two cones. 
The dipole momentum $b_\mu$ denotes the separation of paired Dirac points in the energy-momentum space. More specifically, here $2b_{0}$ denotes the energy separation and one can define the vector $\mathbf{b}=(b_{1},b_{2})=(\mathbf{k}_{D}^{+}-\mathbf{k}_{D}^{-})/2$ denoting the momentum separation, with an example shown in Fig.~\ref{Dirac_cone}(a).

When the system is coupled with an external (effective) electromagnetic field $A_{\mu}$, the effective action gives rise to a topological term,
\begin{equation}
S_{\rm{top}}=i\int d{x}^3\epsilon^{\mu\nu\rho}b_{\mu} \partial_{\nu}A_{\rho}/(2\pi).
\end{equation}
See a derivation in Appendix \ref{AppCurrent}. The response current 
can be obtained by taking the functional derivative of the anisotropic Chern character term,
\begin{equation}
J^{\mu}=\delta S_{\rm{top}}/\delta A_{\mu} =\epsilon^{\mu\nu\rho}\partial_{\nu}b_{\rho}.
\end{equation}
This implies a topological current (see Appendix \ref{AppCurrent})
\begin{equation}
\mathbf{J}_{D}=\frac{1}{2\pi}\left(\nabla b_{0}-\partial_{\tau}\mathbf{b} \right)\times \hat{e}_z, \label{TopCurrent}
\end{equation}
and the corresponding particle density is $\rho_{D}=(\nabla\times\mathbf{b})\cdot\hat{e}_{z}/(2\pi)$, where $\hbar=e=c=1$ is set for briefness and $\tau$ denotes time. Obviously, the topological current $\mathbf{J}_{D}$
is dependent solely on dipole momentum $b_\mu$, which plays a similar role as the gauge field $A_{\mu}\rightarrow\left(b_{0},\mathbf{b}\right)$. The above current stems from a pair of topologically protected Dirac points in 2D, which is different from that caused by the parity anomaly of a single Dirac point (see Appendix \ref{AppCurrent}), and thus may be called  topological currents of dipolar parity anomaly.

In contrast to the conventional currents that are parallel to the induced effective electric field $\mathbf{E}_{\text{eff}}=\nabla b_{0}-\partial_{\tau}\mathbf{b}$, here the pure topological currents are perpendicular to $\mathbf{E}_{\text{eff}}$, similar to the anomalous Hall current arising from a single Dirac point having the corresponding local Berry curvatures.
However, the total anomalous Hall currents contributed from a paired Dirac points will be vanishing in this 2D semimetal system as they carry the opposite Berry curvatures. To the best of our knowledge, this kind of pure topological currents have never been probed experimentally. Although the topological current is derived from an effective field theory of 2D Dirac semimetals under external electromagnetic fields, it can still be illustrated as the particle current in neutral atom systems with the same spatiotemporally dependent Dirac dipole momentum. Notably, when there are multipairs of Dirac points in the system, the total induced topological current equals to that takes the sum of currents arising from each pair of Dirac points \cite{SQC2}. 
The topological current given by Eq. (\ref{TopCurrent}) is general to 2D Dirac semimetals with the low-energy effective Hamiltonian in Eq. (\ref{HamEff}), which can be realized in actual graphene or artificial Dirac systems. In the following section, we propose a 2D Dirac Hamiltonian with the $PT$-symmetry-protected Dirac points, which is realizable with cold atoms in optical lattices and thus enables us to simulate tunable topological currents of dipolar parity anomaly.

\begin{figure}[h!]
\includegraphics[width=1\linewidth]{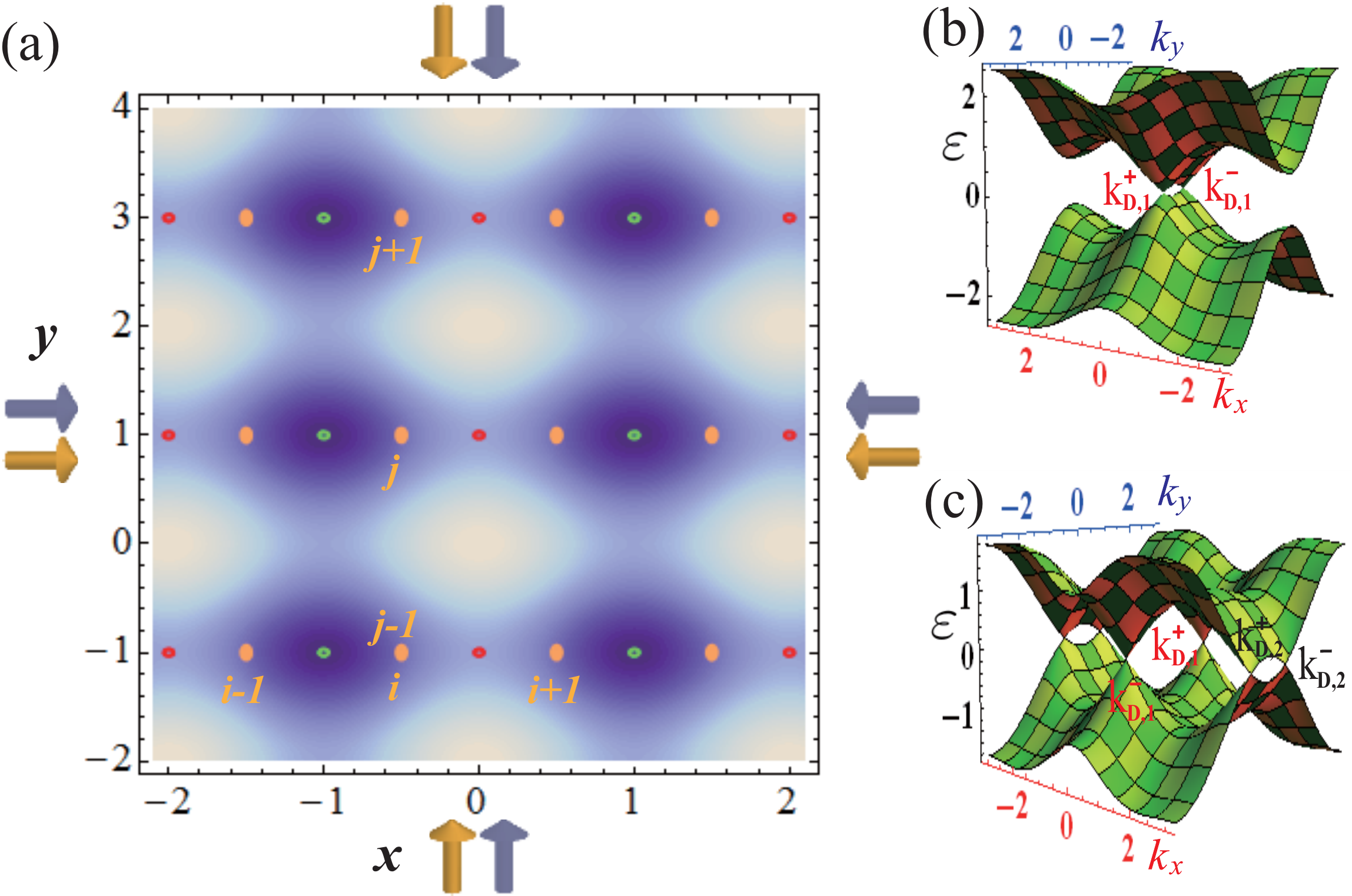}
\caption{(a) A sketch of Raman lattice potential $V^{R}\left(x,y\right)$ as a function of $y$ and $x$ (in units of $a_{x}$ and
let $V^{R}_{0,x}=V^{R}_{0,y}$). The green rings region denote the Raman lattice site [minimal Raman lattice potential $V^{R}_{\rm{Min}}\left(x,y\right)$] and the red rings denote the site with potential $V^{R}\left(x,y\right)=-V^{R}_{\rm{Min}}\left(x,y\right)< V^{R}_{\rm{Max}}\left(x,y\right)$. Moreover, the orange ovals denote the optical lattice site $(x_{i},y_{j})$, where the $i$ and $j$ is the shorthand notation of $x_{i}$ and $y_{j}$, respectively. The orange and light-blue arrows denote the laser beams, which are used to construct the optical lattice and Raman lattice, respectively. From $V^{R}\left(x,y\right)$, we easily find $V^{R}\left(x,y=y_{j}\right)=-V^{R}\left(x+a_{x},y=y_{j}\right)$ as well as $V^{R}\left(x, y\approx y_{j}\right)\approx -V^{R}\left(x+a_{x}, y\approx y_{j}\right)$ and $V^{R}\left(x,y\right)=V^{R}\left(x,y+a_{y}\right)$.
Two typical topological band structures (having one or two pairs of Dirac points) are showed in (b) and (c), where the Dirac points are denoted by $k^{\pm}_{D,1/2}$ and the $\pm$ denote the different chiralities. The parameters are (b), (c) $\delta_{t}=0.32$  and (b), [(c)] $\lambda\left(\tau\right)=1.2$ [$\lambda\left(\tau\right)=0.48$], corresponding to the point of blue (red) lines with $\tau \sim 17.7\,\text{or}\,49.1$ in Fig.~\ref{j-current1}(a).}
\label{Raman-laser}
\end{figure}

\section{Realization in an optical Raman lattice}

We now turn to the realization of the topological particle current $\mathbf{J}_{D}$ with cold atoms in a 2D optical lattice, which can be created by slowly space varying $b_0$ and/or  time varying $\mathbf{b}$, such that the adiabatic approximation is valid for the topological features of the considered bands. For simplicity but without loss of generalization, here we focus on generating $\mathbf{J}_{D}$ from time-dependent $\mathbf{b}$ in the corresponding lattice system, as shown in Fig.~\ref{Dirac_cone}(b). 
In addition, since a pair of Dirac points denoted by $\tau_z$ in the four-component low-energy effective Hamiltonian in Eq. (\ref{HamEff}) is decoupled, we can simply work on a two-component 2D Dirac (graphenelike) Hamiltonian. 
To do this, we consider  the following Bloch Hamiltonian on the 2D lattice:
\begin{eqnarray}
\mathcal{H}\left(\mathbf{k}\right)&=&\sin\left(k_{x}a_{x}\right)\sigma_{1}+\left[\lambda(\tau)-\delta_{t}\cos\left(k_{x}a_{x}\right)\right. \nonumber \\
&&-\cos\left(k_{y}a_{y}\right)\left.\right]\sigma_{2} +f\left(\mathbf{k}\right)\sigma_{0},\label{HK1}
\end{eqnarray}
where $a_{x}$ ($a_{y}$) is the lattice length in the $x$ ($y$) direction, $\lambda(\tau)$ denotes a time-dependent on-site coupling between the two atomic components, $\delta_{t}$ is a hopping parameter, and $f\left(\mathbf{k}\right)$ is an arbitrary function.

According to the theory of joint space-inversion ($P$) and time-reversal ($T$) invariant topological gapless bands \cite{zhao1},
$\mathcal{H}(\mathbf{k})$ supports a nontrivial class of node points with $\mathbb{Z}_{2}$ topological charges. The topological stability of the system only relies on the $PT$ symmetry with the operator $\hat{A}=\hat{P}\hat{T}=\sigma_{1}\hat{K}$, independent of the individual $\hat{P}=\sigma_{1}\hat{I}$ and $\hat{T}=\hat{K}\hat{I}$ here, with $\hat{I}$ inversing the wave vector $k$ to $-k$ and $\hat{K}$ the complex conjugate operator. For $\mathcal{H}(\mathbf{k})$, it is obvious that the individual $P$ and $T$ symmetries are broken, but the joint $PT$ symmetry is preserved. Here we wish to pinpoint that the topological currents arising from any dipole of Dirac points in this lattice Hamiltonian system take the same formula as that for another dipole [i.e., Eq. (\ref{TopCurrent})] derived from the aforementioned continuum model $\mathcal{H}_{\text{eff}}$ because of the topological equivalence between the two dipoles.

The real-space Hamiltonian $H$ corresponding to the Bloch Hamiltonian $\mathcal{H}(\mathbf{k})$ is given by (see Appendix \ref{App_hopping})
\begin{eqnarray}
\!\!\!\!\!H\!\!\!&=&\!\frac{it_x}{2}\sum_{j}\left[\hat{a}_{j+e_{x},\uparrow}^{\dag}\hat{a}_{j,\downarrow}+\hat{a}_{j+e_{x},\downarrow}^{\dag}\hat{a}_{j,\uparrow}\right] \nonumber \\
&&\!+i\frac{\delta_{t}}{2}\sum_{j}\left[\hat{a}_{j+e_{x},\uparrow}^{\dag}\hat{a}_{j,\downarrow}-\hat{a}_{j+e_{x},\downarrow}^{\dag}\hat{a}_{j,\uparrow}\right] \nonumber\\
&&\!+\frac{it_y}{2}\sum_{j}\left[\hat{a}_{j+e_{y},\uparrow}^{\dag}\hat{a}_{j,\downarrow}-\hat{a}_{j+e_{y},\downarrow}^{\dag}\hat{a}_{j,\uparrow}\right] \nonumber \\
&&\!-i\lambda\!\left(\!\tau\!\right)\!\sum_{j}\hat{a}_{j,\uparrow}^{\dag}\hat{a}_{j,\downarrow}
+\!\sum_{j,\sigma,\eta}\!t'_{0,\eta}\hat{a}^{\dag}_{j+e_{\eta},\sigma}\hat{a}_{j,\sigma}+\text{H.c.}, \label{H_r1}
\end{eqnarray}
where $t_x=t_y=t$ [$t\equiv1$ in Eq. (\ref{HK1})] denotes the spin-flip hopping for spin states $\sigma=\uparrow,\downarrow$, and $t'_{0,\eta}$ denotes the spin-independent natural hopping along the $\eta$ axis with $\eta=x,y$. A challenge for realizing this Hamiltonian with neutral atoms is to implement the spin-flip hopping, which acts as the effective 2D spin-orbit coupling. In general, the required spin-flip hopping terms may be achieved in a 2D optical Raman lattice \cite{Liu2014}. Particularly, they can be formed by simultaneously applying two pairs of light beams via two-photon Raman coupling with the lattice potential $V^{L}\left(x,y\right)$ and Raman potential $V^{R}\left(x,y\right)$. A similar Raman lattice scheme was proposed \cite{Liu2014} and realized in the experiment \cite{2D-SOC}.

For realizing our lattice model, we can choose the lattice potential as $$V^{L}\left(x,y\right)=V_{0,x}^{L}\cos^{2}\left(\pi x/a_{x}\right)+V_{0,y}^{L}\cos^{2}\left(\pi y/a_{y}\right),$$ and the Raman potential as $$V^{R}\!\left(x,y\right)\!=i \left\{\!V_{0,x}^{R}\cos\left(\!\pi x/a_{x}\!\right)+\!V_{0,y}^{R}\!\left[\cos\left(\!2\pi y/a_{y}\!\right)+1\right]\!\right\}.$$ The sketch of the Raman potential is shown in Fig.~\ref{Raman-laser}(a) with $a_{y}=2a_{x}$. In addition, the on-site coupling between two atomic components $\lambda(\tau)$ in Eq.~(\ref{H_r1}) can be achieved and tuned by one-photon Raman coupling (via radio-frequency or microwave fields) with the time-dependent Rabi frequency \cite{Gross}. For the optical Raman lattice under tight-binding approximation, the system can be effectively described by the Hamiltonian $H$ in Eq.~(\ref{H_r1}). See Appendix \ref{App_hopping} for more details of the optical Raman lattice and the hopping strengths.

\begin{figure}[h!]
\includegraphics[width=0.8\linewidth]{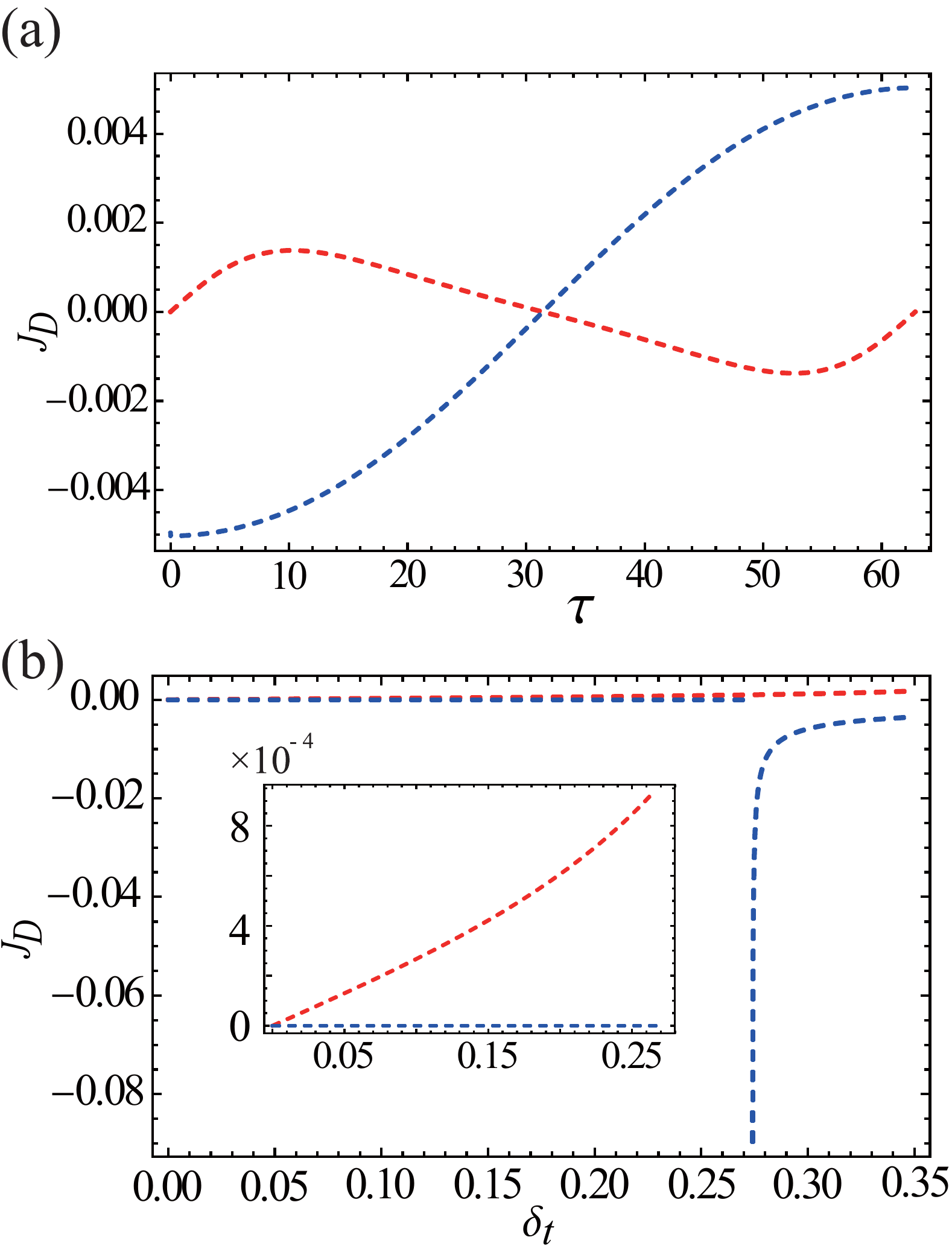}
\caption{The topological current $\mathbf{J}_{D}$ (in units of $t/a_{y}$) for the slowly periodically driving case with $\lambda=0.2$ in (a) and (b). We set $\omega \ll 1$ and $\lambda(\tau)-\lambda_{0}/2\ll 1$ for $\forall \tau$. $\lambda_{0}=2.44,\,1.0$ are denoted by blue and red dashed lines, respectively. In (a), the red and blue lines denote $\mathbf{J}_{D}$ for two pairs and a single pair of Dirac points, respectively, with $\omega=\omega_{\rm{\max}}=0.1$ and $\delta_{t}=0.32$. In (b), $\mathbf{J}_{D}$ as a function $\delta_{t}$ with $\tau=10.0$ is shown. The red line denotes the case with two pairs of Dirac points, while the blue line represents the case undergoing a topological phase transition from the trivial state to the nontrivial state (a pair of Dirac points); the discontinuity of the topological currents $\mathbf{J}_{D}$ reflects this topological phase transition. The inset is an enlarged figure of $\mathbf{J}_{D}$.}
\label{j-current1}
\end{figure}


In this optical lattice, the number of pairs of Dirac points can be controlled by adjusting the values of $\lambda\left(\tau\right)$ and $\delta_t$ (see Appendix \ref{App_Dirac}). If $\lambda \left(\tau\right)>1+\delta_{t}$ or $\lambda \left(\tau\right)<-(1+\delta_{t})$, the band is gapped without Dirac points and thus $\mathbf{J}_{D}=0$. If $ -(1-\delta_{t})<\lambda \left(\tau\right)<1-\delta_{t}$, there are two pairs of Dirac points denoted as $k_{D,1}^{\pm}=\left(0,\pm k_{y,1}\right)$ and $k_{D,2}^{\pm}=\left(\pi,\mp k_{y,2}\right)$, where
$k_{y,1}=\arccos\left[\lambda(\tau)-\delta_{t}\right]/a_{y}$ and
$k_{y,2}=\arccos\left[\lambda(\tau)+\delta_{t}\right]/a_{y}$, respectively.
The corresponding dipole momenta are $2\mathbf{b}_{1}=k_{D,1}^{+}-k_{D,1}^{-}=\left(0,2k_{y,1}\right)$, $2\mathbf{b}_{2}=k_{D,2}^{+}-k_{D,2}^{-}=\left(0,-2k_{y,2}\right)$,
$2b_{0,1}=f\left(k_{D,1}^{+}\right)-f\left(k_{D,1}^{-}\right)$, and $2b_{0,2}=f\left(k_{D,2}^{+}\right)-f\left(k_{D,2}^{-}\right)$.
The topological current can be written as $\mathbf{J}_{D}\!=\mathbf{J}_{D}^{-}+\mathbf{J}_{D}^{+}$, where the two contributions are given by
\begin{equation}
\mathbf{J}_{D}^{\pm}=\mp\lambda^{\prime}(\tau)/\{2\pi a_{y}[1-\!\left(\!\lambda(\tau)\pm\!\delta_{t}\!\right)^{2}]^{1/2}\} \hat{e}_x,
\end{equation}
with $\lambda^{\prime}(\tau)=\partial \lambda/\partial\tau$. Here, we set $b_{0,1/2}=0$ since $f\left(\mathbf{k}\right)$ (even function) only includes the spin-independent hopping.
If $1-\delta_{t}<\lambda \left(\tau\right)< 1+\delta_{t}$ [$-(1+\delta_{t})<\lambda \left(\tau\right)<-(1-\delta_{t})$], there is only one pair of Dirac points $k_{D,1}^{\pm}$ ($k_{D,2}^{\pm}$) and the topological current is $\mathbf{J}_{D}=\mathbf{J}_{D}^{-}$ ($\mathbf{J}_{D}^{+}$).
Two typical bands with one and two pairs of Dirac points are shown in Figs.~\ref{Raman-laser}(b) and~\ref{Raman-laser}(c), respectively. 

For generating  experimentally tunable topological currents, we consider a typical form of $\lambda(\tau)$: the time-periodical driving $\lambda(\tau)=\lambda_P(\tau)=\lambda\cos\left(\omega \tau\right)/2+\lambda_{0}/2$ with frequency $\omega$, 
noting that the concrete form of $\lambda (\tau)$ is not important for generating the topological currents. 
Importantly, the time-dependent term must change slowly enough to avoid breaking the band structure. 
One can set $\lambda/2|\left(\cos(\omega \tau +\omega T_{0})-\cos(\omega\tau)\right)|\ll t$ with $T_{0}=1/t$ being satisfied, such that $\lambda_P(\tau)$ can be considered as a quantity which changes slowly in time. Here $\omega T_{0}=\omega/t\ll1$ is required to avoid resonance absorption for the atom-laser coupling field. Typically, we can set $\omega_{\rm{max}} T_{0}=0.1$ and $\lambda_{\rm{max}}=0.2t$, such that $|\left(\cos(\omega \tau +\omega T_{0})-\cos(\omega\tau)\right)|\leq 0.1$ and $|\lambda(\tau+T_{0})-\lambda(\tau)|_{\rm{max}}=0.01t\ll t$. For the time-periodical driving case, the parameters $\lambda$, $\omega$, and $\tau$ are the compact notations of $\lambda/t$, $\omega T_{0}$, and $\tau/T_{0}$, respectively, which are used to calculate $\mathbf{J}_{D}$. In this notation, $\mathbf{J}_{D}$ takes the unit of $t/a_{y}$. The typical topological currents $\mathbf{J}_{D}$ in this case are shown in Figs.~\ref{j-current1}(a) and~\ref{j-current1}(b). For $\lambda_P(\tau)$, one can regulate the number of the pairs of Dirac points through driving the system. The discontinuity of $\mathbf{J}_{D}$ shown in Fig.~\ref{j-current1}(b) is due to the change of the number of paired Dirac points. This indicates the Lifshitz transition in the system and means that one can induce the transition by adjusting the driving parameters.

\section{Detection schemes}

To observe this topological current in realistic experiments, we consider typical energy scales for several physical quantities. For cold alkali atoms trapped in the optical lattice, the typical recoil energy $E_{R}/(\hbar\times 2\pi)$ is about several $\rm{kHz}$ \cite{Tipical_Temperature}, and the typical hopping amplitude $t$ is changed in the region $(0,0.1E_{R})$. The coherence time of cold atom systems is typically around $\tau_{\rm{coh}}\sim100\rm{ms}$. Thus, the intrinsic timescale of the lattice system is $T_{0}=\hbar/t=(1, 10,100,1000)\rm{ms}$, corresponding to $t=(1000,100,10,1)\rm{Hz}$.  The bosonic or fermionic atoms can be loaded in the optical Raman lattice, and we assume the bosons as used in the experiment \cite{2D-SOC}. The critical temperature of typical Bose gases is about $10-100\rm{nK}$. 
When the trapping potential satisfies $V^{L}_{0,x/y}\gtrsim 5 E^{x/y}_{R}$,
the system is well described within the tight-binding approximation \cite{Tipical_Temperature}. The bandwidth $W\approx 4\left(t_{0,x}^{\prime}+t_{0,y}^{\prime}+t_x+t_y\right)$ is several $\rm{kHz}$ (the corresponding temperature is several-hundred $\rm{nK}$) for the 2D systems, where $t^{\prime}_{0,x/y}$ is the amplitude of normal hopping. Thus, if the temperature of the system is about several hundred $\rm{nK}$ (the same level of bandwidth $W$), we can assume the bosonic atoms will be an incoherent homogeneous distribution within each band in $\mathbf{k}$ space \cite{Measuring-chern-number} (each $\mathbf{k}$ can be equally filled). Under these conditions, the topological current can be detected via measuring the drift of the center of mass of the atomic cloud along the $x$ direction, although it is not straightforward to perform conventional transport measurements in cold-atom systems.

\begin{figure}[h!]
\includegraphics[width=0.8\linewidth]{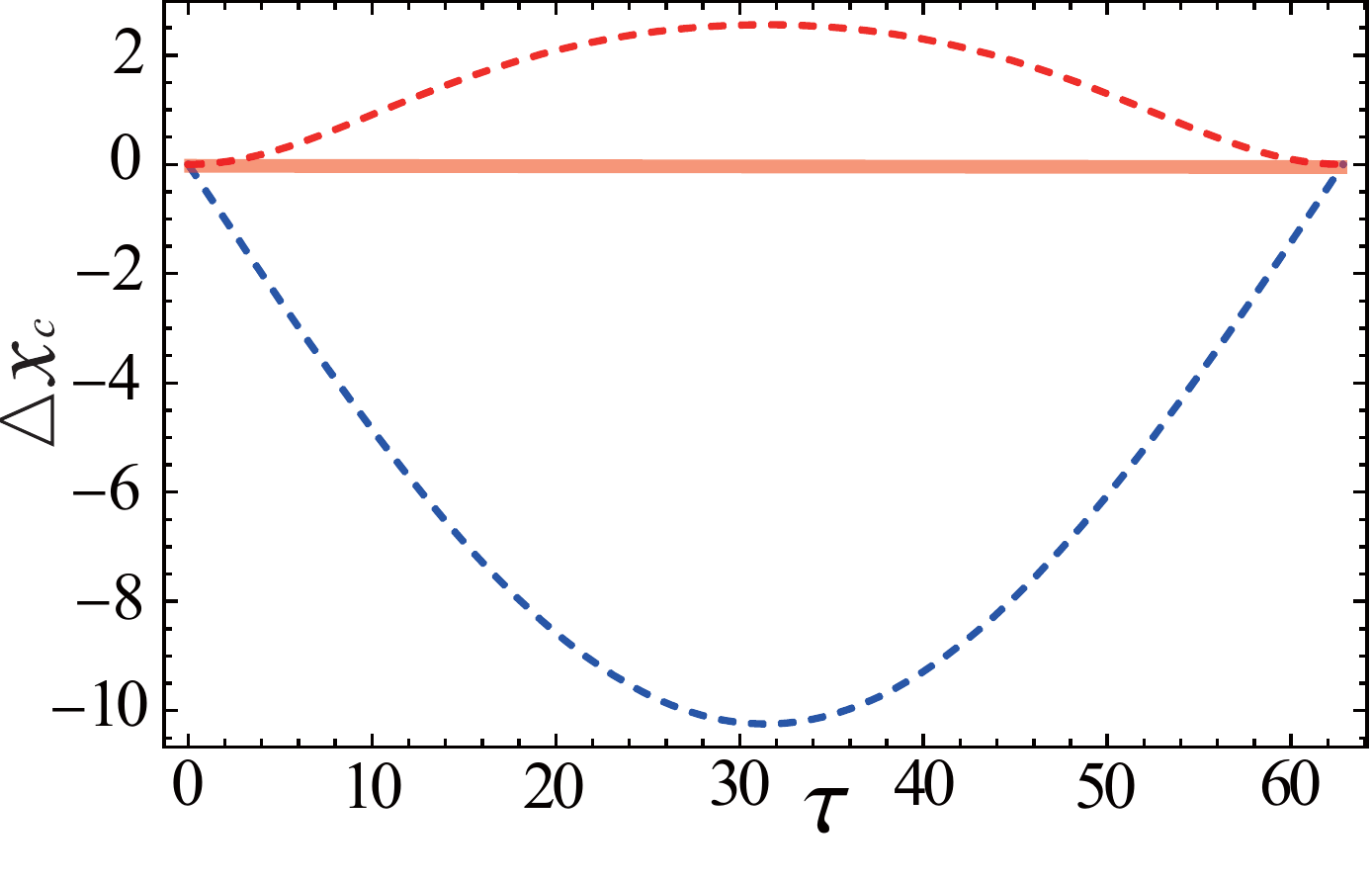}
\caption{The drift of the center of mass $\Delta x_{c}=x_{c}(\tau)-x_{c}(\tau=0)$ (in units of $a_{x}$) as a function of $\tau$ in the $x$ direction are shown by taking $\rho=0.01(a_{y}a_{x})^{-1}$. The parameters for blue and red dashed lines are the same as those in Fig.~\ref{j-current1}(a). Here the parameters for the red transparent thick line are the same as the others color lines except $\delta_{t}$, where $\delta_{t}$ is $0$ ($0.32$) for the red transparent thick line (the others lines). It is clear that there is no drift of the center of mass $\Delta x_{c}$ for $\delta_{t}=0$, even though there are two pairs of Dirac points (due to the current cancellation from the two pairs). Thus this nonvanishing $\Delta x_{c}$ feature may be taken as an experimental signature for topological currents of dipolar parity anomaly.}
\label{centel-of-mass}
\end{figure}

The drift measurement can be achieved in the same way as that in the experiment in Ref. \cite{Measuring-chern-number}, but without an external effective force since the topological currents arise from the time-varying fields here. The drift velocity reads $v_{c}=\mathbf{J}_{D}/\rho$, where $\rho$ denotes the average density of the atomic gas and is assumed invariable within the coherence time. Thus the atomic center-of-mass displacement reads $x_{c}(\tau)=\int \mathbf{J}_{D} d\tau/\rho$, which directly reveals the features of $\mathbf{J}_{D}$. The typical results of the drift of the center of mass $\Delta x_{c}=x_{c}(\tau)-x_{c}(\tau=0)$ as a function of time $\tau$ are shown in Fig.~\ref{centel-of-mass} for the periodical driving case with $\rho=0.01 (a_{y}a_{x})^{-1}$ [see Fig.~\ref{j-current1}(a) for the corresponding $\mathbf{J}_{D}$]. In Fig.~\ref{centel-of-mass}(a), the dashed blue lines indicate the maximum drift of $|\Delta x_{c}|$. To detect $\mathbf{J}_{D}$ for the hopping amplitude $t\sim1\rm{kHz}$, one requires the particle density $\rho<0.1\,(a_{y}a_{x})^{-1}$. This can be achieved by setting the particle number $N\lesssim1000$ for the typical optical lattice of size $100 a_{x}\times 100 a_{y}$. For $\rho=0.1,\, 0.05, 0.01 (a_{y}a_{x})^{-1}$, the corresponding maximum displacement of $|\Delta x_{c}|$ is about $1a_{x},\, 2a_{x}, \, 10a_{x}$ with $\tau \sim 31.4\rm{ms}$, which can be well measured in cold-atom experiments \cite{Measuring-chern-number}.

An alternative detection method is to use an atomic gas of two wave packets prepared near the paired Dirac points and then to observe the topological response from the dipolar parity anomaly, via the same dynamical (\emph{in situ}) density measurements \cite{Measuring-chern-number}. For the considered time-varying perturbation that is uniform in space, the semiclassical dynamics of the wave packets that are centered at $\mathbf{r}_c$ (real space) and $\mathbf{k}_c$ (momentum space) can be described by the equations of motion \cite{Xiao2010,axial_GF}: $\dot{\mathbf{r}}_c=\nabla_{\mathbf{k}}\varepsilon_{\mathbf{k}}-\Omega_{\mathbf{kk}}\dot{\mathbf{k}}_c-\Omega_{\mathbf{kt}}+\mathbf{E}_{\text{eff}}\times \hat{e}_z$ and $\dot{\mathbf{k}}_c=0$, where $\varepsilon_{\mathbf{k}}$ is the energy dispersion, and $\Omega_{\mathbf{kk}}$ and $\Omega_{\mathbf{kt}}$ are (generalized) Berry curvatures. Since the Berry curvatures are opposite near a pair of Dirac points, the terms with respect to $\Omega_{\mathbf{kk}}$ and $\Omega_{\mathbf{kt}}$ have vanishing total contributions in the real-space motion of the atomic gas. Finally, the contribution from the term $\mathbf{E}_{\text{eff}}\times \hat{e}_z$ as the topological response can be extracted (isolated) by factoring out the effect of the group velocity $\nabla_{\mathbf{k}}\varepsilon_{\mathbf{k}}$ in the dynamics. This can be achieved by simply subtracting the responses of two protocols with opposite $\mathbf{E}_{\text{eff}}$ (i.e., $\partial_{\tau}\mathbf{b}$ in this case), similar to the differential measurements of atomic center-of-mass shifts in Ref. \cite{Measuring-chern-number}.

\section{Discussion and Conclusion}

Before concluding, we note that the topological current $\mathbf{J}_{D}$ is protected by the joint $PT$ symmetry, which can also be tested in experiments. If we add the $P$ and/or $T$ broken perturbation term, such as $\mathcal{H}^\prime=\epsilon\sin(k_{y}a_{y})$ with a small value $\epsilon$, $\mathbf{J}_{D}$ is not destroyed. If we add the $PT$-symmetry-broken term, such as $ \epsilon \sigma_{z}$ in this system, the bands will be gapped and then $\mathbf{J}_{D}$ will disappear.

In a realistic cold-atom system with an external harmonic trap, one should consider the mixture of the propagating edge modes and the bulk currents \cite{PNAS2013}, if they have contributions along the same axis. In our proposed cold-atom system, the edge currents in the 2D system can only propagate along the particular direction of an effective external force, but the intrinsic pure topological currents in the bulk can be tuned to be perpendicular and then be isolated from the edge currents. For the case of a paired of Dirac points shown in Fig.~\ref{Dirac_cone}, the possible edge currents can only propagate in the metallic edges along the $y$ axis, while the topological currents in the bulk are along the $x$ axis. Furthermore, in the center-of-mass measurements, the wave packet of an atomic gas can be prepared in the center of the 2D optical lattice in the presence of the harmonic trap, such that the edge effects can be neglected in the dynamical response when the wave packet does not reach the trap edge. As shown in Fig.~\ref{centel-of-mass}, the length scale of atomic drifts is about several lattice sites within the timescale about 30 ms, which can be achieved in realistic experiments.

In summary, we have revealed an exotic kind of topological currents arising from the dipolar parity anomaly in the presence of the spatiotemporally weak-dependent energy-momentum separation of paired Dirac points in 2D $PT$-symmetric semimetals. In particular, we have proposed an experimentally feasible scheme to realize and detect this kind of topological particle currents with ultracold atoms in a 2D optical lattice. The present scheme is quite promising for realizing the first experimental detection of topological currents of dipolar parity anomaly.

\begin{acknowledgments}
We thank Y. X. Zhao for helpful discussions. This work was supported by the NKRDP of China (Grant No. 2016YFA0301800), the NSFC (Grants No. 11604103 and No. 11474153), the NSAF (Grant No. U1830111), the NSF of Guangdong Province (Grant No. 2016A030313436), the KPST of Guangzhou (Grant No. 201804020055), the Startup Foundation of Anhui University (Grant No. J01003310), the Startup Foundation of SCNU, and the GRF (Grants No. HKU 173309/16P and No. HKU173057/17P) and a CRF (Grant No. C6005-17G) of Hong Kong.
\end{acknowledgments}

\begin{appendix}

\section{Topological Currents arising from dipolar parity anomaly }\label{AppCurrent}
This novel topological current $\mathbf{J}_{D}$ can be caused by the action $S=i\int dx^{3}\epsilon^{\mu\nu\rho}b_{\mu}\partial_{\nu}A_{\rho}/2\pi$, where $A_{\mu}$ represents the corresponding component of the electric-magnetic potential, and $\epsilon^{\mu\nu\rho}$ is the antisymmetric tensor. The details of how to obtain the topological current $\mathbf{J}_{D}$ are presented below. One
may start with the graphenelike effective Hamiltonian that reads $\mathcal{H}_{\text{eff}}=k_{\mu}\gamma^{\mu}-b_{\mu}\gamma^{\mu}\tau^{z}$.
 When the system is coupled with an external electromagnetic filed $A_{\mu}$, the effective action $S_{\rm{eff}}$ can be written as
\begin{equation}
S_{\rm{eff}}=\ln \rm{det}(i\, /\kern-0.7em D -/\kern-0.5em b\tau^{z}).
\end{equation}
Here $D$ is the gauge covariant derivative: $D_\mu = \partial_\mu + A_\mu$,
and the symbol ``$/$" denotes the inner product of the vector and the Gamma matrices (namely, $/\kern-0.5em{b} = b_\mu \gamma^\mu$) in the (2+1) dimension of space-time.  We first regularize the effective action through the Pauli-Villars method,
\begin{equation}
S_{\rm{eff}}^{\rm{Reg}}=\rm{tr}\ln \rm{det}(i\, /\kern-0.7em D -/\kern-0.5em b\tau^{z}+m \sigma_{z}),
\end{equation}
where $m \sigma_{z}$ is the regularization mass term. After a straightforward expansion, we can obtain
\begin{eqnarray}
S_{\rm{eff}}\left[A,M\right]\!\!\!&=&\!\!\!\rm{tr \ln}\left(i /\kern-0.5em \partial +m \sigma_{z} \right)+\rm{tr}\left(\frac{1}{i /\kern-0.5em\partial +m \sigma_{z}}\left(/\kern-0.7em A-/\kern-0.5em b\tau^{z}\right)\right)\nonumber \\
&&\!\!\!+\frac{1}{2}\rm{tr}\!\left(\frac{1}{i /\kern-0.5em\partial +\!m \sigma_{z}}\left(/\kern-0.7em A-/\kern-0.5em b\tau^{z}\right)\right.\nonumber \\
&&\!\!\! \times \left.\frac{1}{i /\kern-0.5em\partial +\!\rm{m} \sigma_{z}}\left(/\kern-0.7em \rm{A}-/\kern-0.5em \rm{b}\tau^{z}\right)\!\right)+\cdots.
\end{eqnarray}

Also, what concerns us is only the quadratic term. Accordingly, we restrict our attention
to the  terms $b\cdot \Pi\cdot A$ and $A\cdot \Pi\cdot b$, which contribute to effective action equally. Since only the vacuum polarization graph and the triangle graph are ultraviolet divergent, the only term of concern should be the third one, which is quadratic in the gauge field $A$.
The third term gives
\begin{equation}
S_{\rm{top}}=\int \frac{d^{3}p}{(2\pi)^{3}}b_{\mu}(-p)\Pi^{\mu\nu}(p) A_{\nu}(p),
\end{equation}
and
\begin{eqnarray}
\Pi^{\mu\nu}(p)&=&4m\epsilon^{\mu\nu\rho}p_{\rho}\int \frac{d^{3}p}{(2\pi)^{3}}\frac{1}{((p+q)^{2}+m^{2})(q^{2}+m^{2})}\nonumber \\
&\sim&\frac{1}{2\pi}\rm{sgn}(m)\epsilon^{\mu\nu\rho}p_{\rho}.
\end{eqnarray}
After a straightforward, but lengthy and tedious calculation, the anisotropic term is finally obtained
\begin{equation}
S_{\rm{top}}=\pm\frac{i}{2\pi}\int d^{3}x\epsilon^{\mu\nu\rho}b_{\mu} \partial_{\nu}A_{\rho}.
\end{equation}
Since the sign does no harm in the effective action, one arrives at a compact form,
\begin{equation}
S_{\rm{top}}=\frac{i}{2\pi}\int d^{3}x\epsilon^{\mu\nu\rho}b_{\mu} \partial_{\nu}A_{\rho} \label{Seff}
\end{equation}
Let us consider the response under an external electromagnetic field $A$. The response current
of topological semimetals can be easily obtained by taking the functional derivative of the
anisotropic Chern character term [as shown in Eq.~(\ref{Seff})] with respect to the electromagnetic
field $A$. Accordingly, one finds
\begin{equation}
J^{\mu}=\frac{\delta}{\delta A_{\mu}} S_{\rm{top}} =\frac{1}{2\pi}\epsilon^{\mu\nu\rho}\partial_{\nu}b_{\rho},
\end{equation}
which implies
\begin{eqnarray}
\rho_{D}&=&\frac{1}{2\pi}(\nabla\times\mathbf{b})\cdot\hat{e}_{z}, \\
J_{D}&=&\frac{1}{2\pi}\left(\nabla b_{0}-\partial_{t}\mathbf{b} \right)\times \hat{e}_z \label{J_D}
\end{eqnarray}
where $\rho_{D}$ is the density and the corresponding current is $J_{D}$. Here, the first term in $J_{D}$ depends upon the spatial gradient of
difference in energy between two Dirac points, while the second one depends on the time
derivative of the span in momentum space between two Dirac points. In particular, it should
be noted that there is no dependency on the original gauge field $A$.

Meanwhile, as a comparison, one could evaluate the response current of a single Dirac
point coupled to an electromagnetic field in a similar manner. The functional derivative of
the Chern-Simons term yields
\begin{equation}
J^{\mu}=\frac{\delta}{\delta A_{\mu}} S_{\rm{CS}} =\frac{1}{2\pi}\epsilon^{\mu\nu\rho}\partial_{\nu}A_{\rho}=\frac{F}{2\pi},
\end{equation}
which implies
\begin{eqnarray}
\rho&=&\frac{1}{2\pi}(\nabla\times\mathbf{A})\cdot\hat{e}_{z}=\frac{B}{2\pi}, \\
J&=&\frac{1}{2\pi}\left(\nabla A_{0}-\partial_{t}\mathbf{A} \right)\times \hat{e}_z =\frac{1}{2\pi}\mathbf{E}\times \hat{e}_z, \label{J_CS}
\end{eqnarray}
where $B$ and $\mathbf{E}$ denote the magnetic field ($z$ component) and electric field, respectively. The
formulas above indicate that a Chern-Simons term would induce a transverse conductivity
in even dimensions. In topological semimetals, as a consequence of the emergence of the anisotropic
Chern character term, a new sort of response, where the span of two Dirac points $b_{\mu}$ plays
the role of the gauge field, emerges, as shown in Eq.~(\ref{J_D}).

\section{The atomic spin-flip-hopping in the 2D optical Raman lattice}\label{App_hopping}
\begin{figure}[h]
\includegraphics[width=0.85\linewidth]{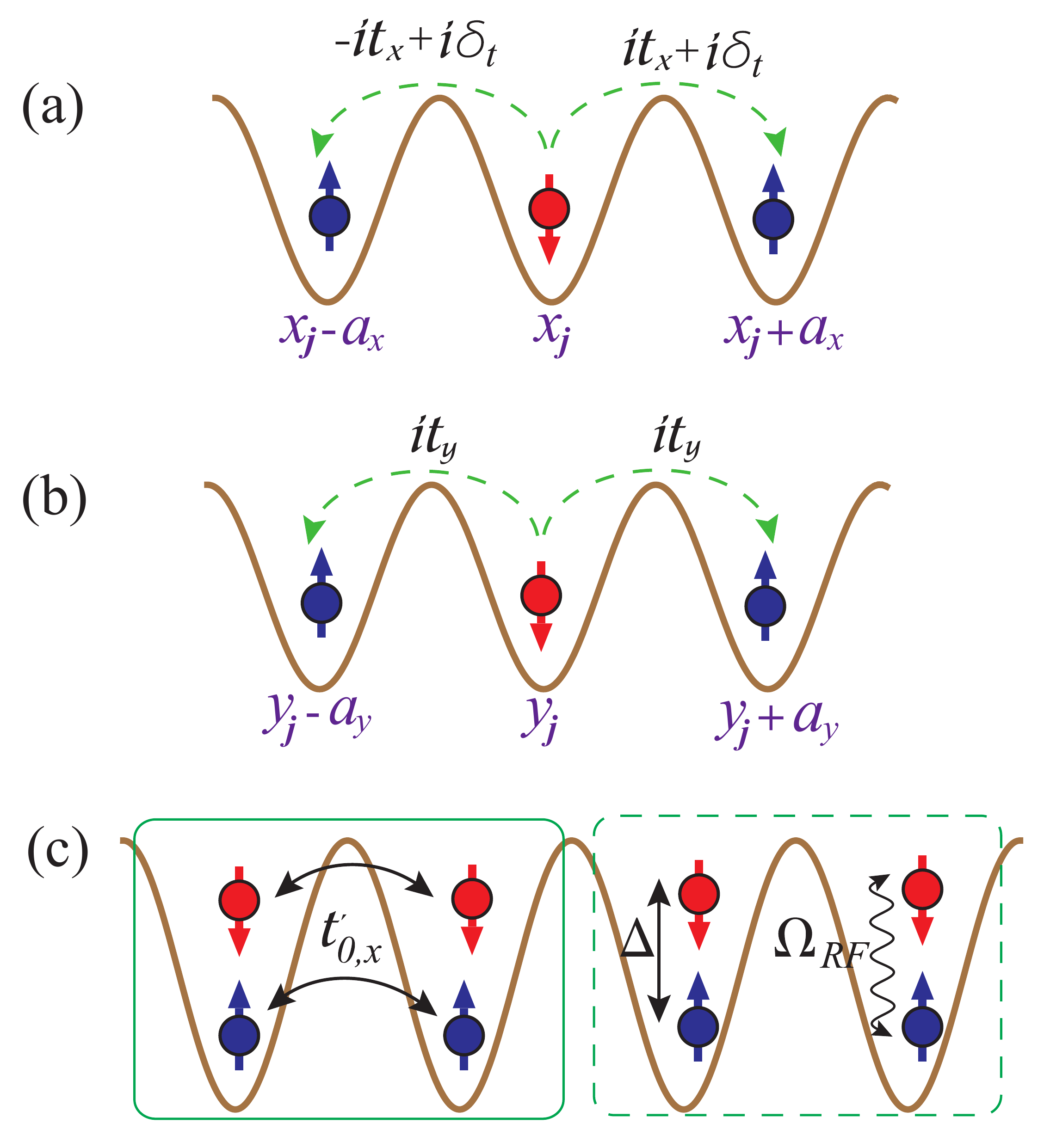}
\caption{(a)--(c) The indicate the cartoon pictures of the spin-flip hopping along the $x$ and $y$  directions, the radio-frequency field induced on-site spin-flip hopping, and the normal spin-independent hopping, respectively, where $x_{j}$ ($y_{j}$) denotes the lattice site along the $x$ ($y$) direction and $a_{x}$ ($a_{y}$) denotes the lattice length along the $x$ ($y$) direction.  $t^{\prime}_{0,x}$  is the amplitude of normal spin-independent hopping along the $x$ direction (we do not show the $y$ direction), and $\Delta$  is the detuning of the $|\uparrow\rangle$ state and $|\downarrow\rangle$. }
\label{catoon}
\end{figure}

If we can realize the real-space Hamiltonian as
\begin{eqnarray}
\!\!\!\!\!\widetilde{H}\!\!\!&=&\!it_{x}\sum_{j}\left[\hat{a}_{j+e_{x},\uparrow}^{\dag}\hat{a}_{j,\downarrow}+\hat{a}_{j+e_{x},\downarrow}^{\dag}\hat{a}_{j,\uparrow}\right] \nonumber \\
&&\!+i\delta_{t}\sum_{j}\left[\hat{a}_{j+e_{x},\uparrow}^{\dag}\hat{a}_{j,\downarrow}-\hat{a}_{j+e_{x},\downarrow}^{\dag}\hat{a}_{j,\uparrow}\right] \nonumber\\
&&\!+it_{y}\sum_{j}\left[\hat{a}_{j+e_{y},\uparrow}^{\dag}\hat{a}_{j,\downarrow}-\hat{a}_{j+e_{y},\downarrow}^{\dag}\hat{a}_{j,\uparrow}\right] \nonumber \\
&&\!-i\lambda\!\left(\!\tau\!\right)\!\sum_{j}\hat{a}_{j,\uparrow}^{\dag}\hat{a}_{j,\downarrow}
+\!\sum_{j,\sigma,\boldsymbol{\eta}}\!t^{\prime}_{\sigma}\hat{a}^{\dag}_{j+\!\boldsymbol{\eta},\sigma}\hat{a}_{j,\sigma}+\!H.c.,\label{H_r}
\end{eqnarray}
where  $\boldsymbol{\eta}=\left(e_{x}, e_{y}\right)$, $\sigma=\uparrow,\downarrow$ and $\lambda\left(\tau\right)=\Omega_{\rm{RF}} \cos\left(\omega\tau\right)+\Omega_{0}$ for periodical drive, then the nondimensionalized real-space Hamiltonian $H$  in Eq. (3) is realized by doing a transformation $\widetilde{H}/(2t_{x})\rightarrow H$  with $t_{x}=t_{y}=1$. The processes for spin-flip hopping in the $x$ ($y$) direction [the hopping process from the lattice site $x_{j}$ ($y_{j}$) to site $x_{j}\pm a_{x}$ ($y_{j}\pm a_{y}$)], the radio-frequency field induced on-site spin-flip hopping and normal spin-independent hopping are presented in Fig.~\ref{catoon}.
We only need to realize the processes (a)--(c) in cold-atom systems; then the momentum space $\mathcal{H}\left(\mathbf{k}\right)$ is simulated since the processes of $ \downarrow \leftarrow \uparrow\rightarrow \downarrow$ are only the Hermitian processes of  $\sum_{j}$  (a) and (b).

\begin{figure}[h!]
\includegraphics[width=0.8\linewidth]{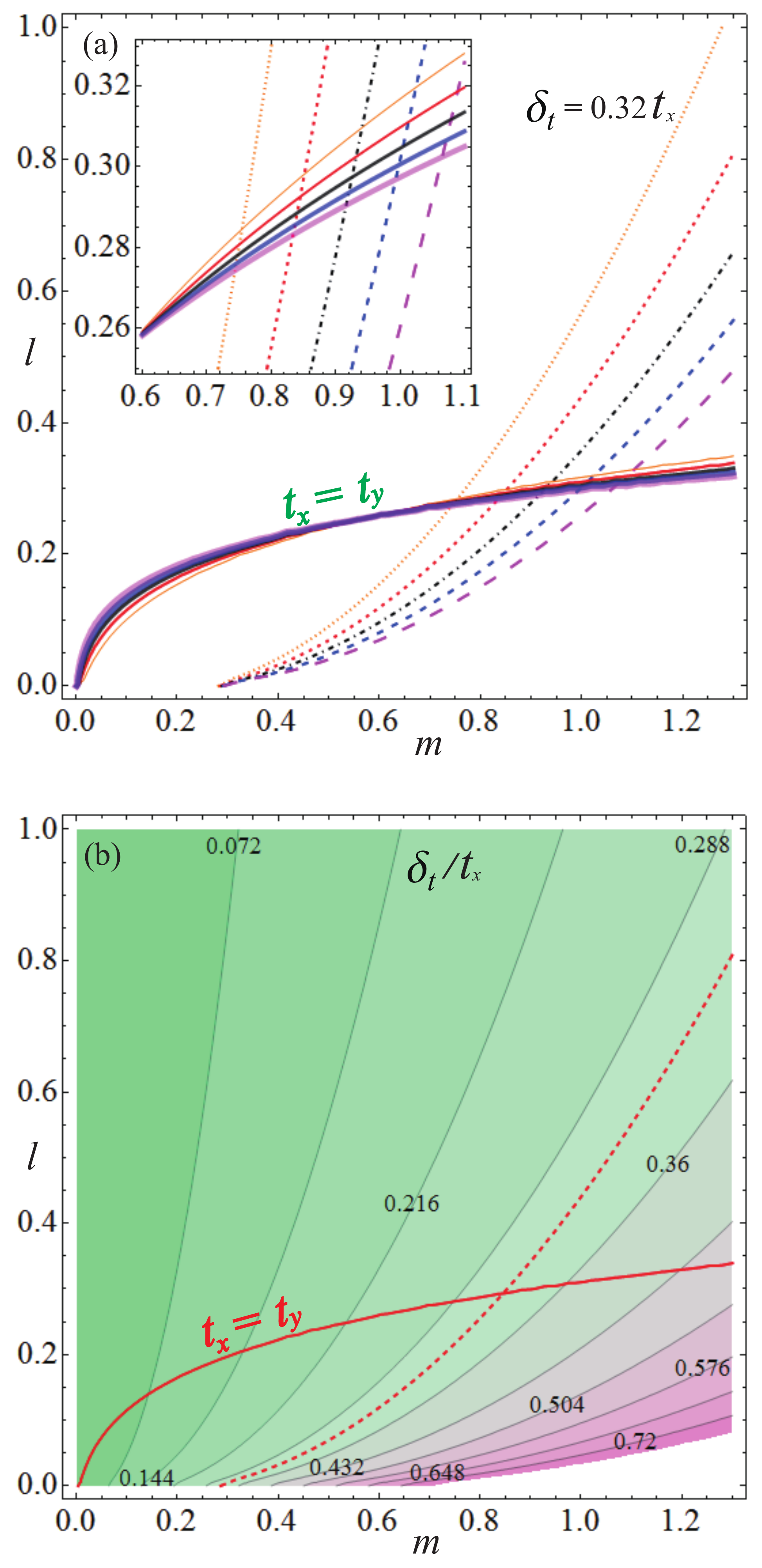}
\caption{ (a) The solid lines denote $t_{x}=t_{y}$ and the dashed lines denote the $\delta_{t}=0.32t_{x}$, which have been chosen to calculate the topological current $J_{\rm{D}}$ and the corresponding drifting of center mass $x_{\rm{c}}$; the orange, red, black, blue, and magenta lines denote the values of trapping potential $\widetilde{V}_{0,x}^{L}=4,\,5,\,6,\,7,\,8$, respectively. (b) $\delta_{t}/t_x$ as a function of $m$ and $l$ with $\widetilde{V}_{0,x}^{L}=5$.}
\label{tx_delta_app}
\end{figure}

The process (a) with $\delta_{t}=0$ requires that the $\Omega_{eff}\left(x\right)$ satisfy the constraint as $V^{R}\left(x ,y\right)= -V^{R}\left(x+a_{x},y \right)$. When we make the Raman lattice length meet the condition $a^{R}_{x} =2a_{x}$, this requirement can be satisfied, where $a_{x}$ is the lattice length of trapping the optical lattice in the $x$ direction. The process (b) requires that the $\Omega_{eff}\left(y\right)$ satisfy the constraint as $V^{R}\left(x ,y\right)= V^{R}\left(x,y+a_{y}\right)$. If the Raman lattice length $a^{R}_{y} $ is equal to $a_{y}$, the process (b) can be realized. If we can independently control the Raman lattice and trapping lattice respectively, we can simulate this momentum space $\mathcal{H}\left(\mathbf{k}\right)$ by choosing Raman potential $V^{R}\left(x,y\right)$ as
\begin{equation}
V^{R}\!\left(x,y\right)\!=i \left\{\!V_{0,x}^{R}\cos\left(\!\frac{\pi x}{a_{x}}\!\right)+V_{0,y}^{R}\!\left[\!\cos\left(\!\frac{2\pi y}{a_{y}}\!\right)+1\right]\!\right\},\label{V_R}
\end{equation}
and trapping lattice potential $V^{L}\left(x,y\right)$ as
\begin{equation}
V^{L}\left(x,y\right)=V_{0,x}^{L}\cos^{2}\left(\frac{\pi x}{a_{x}}\right)+V_{0,y}^{L}\cos^{2}\left(\frac{\pi y}{a_{y}}\right).\label{V_L}
\end{equation}
Here the imaginary unit $i$ in Eq.~(\ref{V_R}) can be realized by suitable tuning of the relative phase of the Raman frequency $\Omega_{1}\Omega_{2}^{\ast}$.
From Eqs.~(\ref{V_R}) and (\ref{V_L}), we can easily obtain Raman lattice length $a_{x}^{R}=2a_{x}$ ($a_{y}^{R}=a_{y}$) in the $x$ ($y$) direction. In order to  realize the Raman potential easily, we can choose a square lattice as the Raman potential, i.e., $a_{y}=2a_{x}$. The lattice site $x_{i}$ can be written as  $x_{i}=\left(1/2+i\right)a_{x}$ and $y_{j}$ can be written as  $y_{j}=\left(1/2+j\right)a_{y}$. The Raman potential $V^{R}\left(x,y\right)$ as a function of $y$ and $x$ is presented in Fig. 2(a). When $y=y_{j}$, this Raman potential $V^{R}\left(x,y\right)$  can easily meet the request $V^{R}(x ,y)= -V^{R}(x+a_{x},y)$ (green (red) rings in Fig. 2(a)  when we choose $x=(x_{i-1}+x_{i})/2$ [$x+a_{x}=(x_{i}+x_{i+1})/2$]), which is the condition for realizing the process (a) with $\delta_{t}=0$. Without loss of generality, we choose the value of $y$  approaching $y_{j}$ as an example to analyze how can we realize the process (a) with nonzero $\delta_{t}$. If $y$ is equal to $y_{j}\pm\delta$, process (a) with $\delta_{t}=0$ is not quantitatively satisfied since the value of $V^{R}\left[x\in\left(x_{i-1},x_{i}\right),y=y_{j}\pm\delta\right]$ is not strictly equal to $-V^{R}\left[x+a_{x},y=y_{j}\pm\delta\right]$ when $\delta\neq 0$. This small deviation of $V^{R}\left[x\in\left(x_{i-1},x_{i}\right),y=y_{j}\pm\delta\right]$ with $-V^{R}\left[x+a_{x},y=y_{j}\pm\delta\right]$ has a significant contribution for the realization of nonzero $\delta_{t}$. On the other hand, approaching the lattice $x_{i}$, the Raman potential in the $x$ direction is antisymmetrical ($\cos\left[\pi( x_{i}+\delta)/a_{x}\right]=-\cos\left[\pi (x_{i}-\delta)/a_{x}\right]$), whereas the Wannier functions are symmetrical for  $x=x_{i}\pm \delta$, and thus the contribution of the Raman potential $V_{0,x}^{R}\cos\left(\!\pi x/a_{x}\!\right)$ in the $y$-direction hopping is zero. This means that the hopping amplitude in the $y$ direction is independent of the site $x_{i}$. The process (b) is always satisfied for all value of $y$ since the Raman potential satisfies $V^{R}\left(x,y\right)=V^{R}\left(x,y\pm a_{y}\right)$ and the hopping amplitude in the $y$ direction is independent of site $y_{i}$. Thus, if we can realize those specific  Raman and lattice potentials, which are presented in Eqs.(\ref{V_R}) and (\ref{V_L}), the real-space Hamiltonian $H$ can be realized. In consequence, it will cause the existence of these topological currents of dipolar parity anomaly $\mathbf{J}_{D}$ in the 2D optical lattice.

\begin{figure}[h!]
\includegraphics[width=0.8\linewidth]{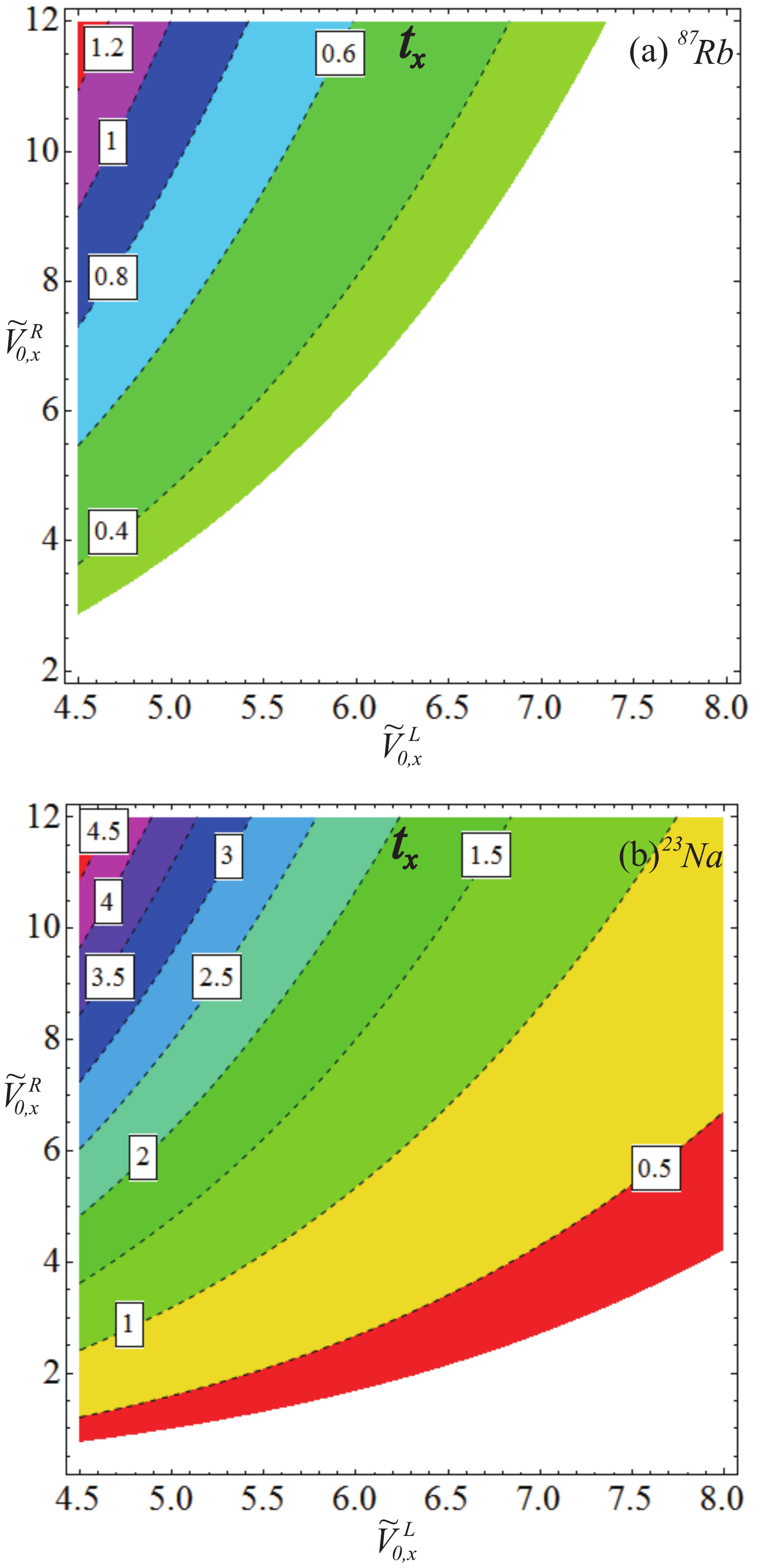}
\caption{$t_{x}$ as a function of trapping potential $\widetilde{V}_{0,x}^{L}$ and Raman potential $\widetilde{V}_{0,x}^{R}$ takes the unit of $\rm{kHz}$ for (a) $^{87}\rm{Rb}$ and (b) $^{23}\rm{Na}$. In these regions, the maximum drifting of the center of mass $\Delta x_{c}$, which is presented in Fig.~4, will be observed within the coherent time $\tau_{\rm{coh}}$. The trapping laser wavelength is $\lambda=790.1\rm{nm}$ and recoil energy is $E_{R}^{x}=(k_{x})^{2}\hbar^{2}/(2M)$ ($E_{R}^{x}=\hbar\times 2\pi\times 3.68\rm{kHz}$ for $^{87}\rm{Rb}$ and $E_{R}^{x}=\hbar\times 2\pi\times 13.9 \rm{kHz}$ for $^{23}\rm{Na}$), where $k_{x}=2\pi/\lambda$ is the recoil momentum \cite{tunable-so}.}
\label{tx_ty_app}
\end{figure}

In order to calculate the hopping amplitude qualitatively, we can use the harmonic approximation at the minimum
trapping potential point  \cite{jiang-pra-2,Wannier}, and thus the Wannier function in the $x$ and $y$ directions can be written as
\begin{equation}
w(x-x_{i})=\left(\widetilde{V}_{0,x}^{L}\right)^{\frac{1}{8}}\!\left(\frac{\pi}{a_{x}^{2}}\right)^{\frac{1}{4}}\!
\exp\left[\!-\frac{\pi^{2}}{2}\sqrt{\widetilde{V}_{0,x}^{L}}\frac{\left(x-x_{i}\right)^{2}}{a_{x}^{2}}\!\right],
\end{equation}
and
\begin{equation}
w(y-y_{i})=\left(\widetilde{V}_{0,y}^{L}\right)^{\frac{1}{8}}\!\left(\frac{\pi}{a_{y}^{2}}\right)^{\frac{1}{4}}\!
\exp\left[\!-\frac{\pi^{2}}{2}\sqrt{\widetilde{V}_{0,y}^{L}}\frac{\left(y-y_{i}\right)^{2}}{a_{y}^{2}}\!\right],
\end{equation}
respectively, where $\widetilde{V}_{0,x/y}^{L}=V^{L}_{0,x/y}/E_{R}^{x/y}$ is the dimensionless optical lattice depth in
unit of $E_{R}^{x/y}$ and $E_{R}^{x/y} = \hbar^{2} k_{x/y}^2/2m (k_{x/y}=2\pi/\lambda_{x/y}, \lambda_{y}=2\lambda_{x}=4a_{x})$ is the
recoil energy. The spin-flip hopping term induced by the Raman field takes the form $t_{(i,j),(\downarrow,\uparrow)}=\int w^{\ast}_{\uparrow}\left(\mathbf{r}-\mathbf{r}_{i}\right)V^{R}\left(x,y\right)w_{\downarrow}\left(\mathbf{r}-\mathbf{r}_{j}\right)d\mathbf{r}$. After some lengthy but straightforward calculations, the explicit form of the spin-flip hopping term in the $x$ and $y$ directions can be written as $t_{(i,j),(\downarrow,\uparrow)}^{x}=\pm it_{x}+i\delta_{t}$ (for $j=i\pm1$) and $t_{(i,j),(\downarrow,\uparrow)}^{y}= it_{y}$,
where $\pm$ is corresponding to $j=i+1$ and $j=i-1$.
Here the forms of $t_{x}$, $\delta_{t}$, and $t_{y}$  are written as
\begin{eqnarray}
t_{x}=V_{0,x}^{R}\exp{\left[-\frac{\pi^{2}\sqrt{\widetilde{V}_{0,x}^{L}}}{4}\right]}
\exp{\left[-\frac{1}{4\sqrt{\widetilde{V}_{0,x}^{L}}}\right]},
\end{eqnarray}
\begin{eqnarray}
\delta_{t}=V_{0,y}^{R}\!\exp{\!\left[\!-\frac{\pi^{2}\sqrt{\widetilde{V}_{0,x}^{L}}}{4}\right]}
\!\!\left[1-\!\exp{\!\left[\!-\frac{1}{\sqrt{\widetilde{V}_{0,y}^{L}}}\right]}\!\right],
\end{eqnarray}
\begin{equation}
t_{y}=V_{0,y}^{R}\!\exp{\!\left[\!-\frac{\pi^{2}\sqrt{\widetilde{V}_{0,y}^{L}}}{4}\right]}
\left[1+\!\exp{\!\left[-\frac{1}{\sqrt{\widetilde{V}_{0,y}^{L}}}\right]}\right].
\end{equation}
 By taking the units of $V_{0,x}^{R}$ and $V_{0,x}^{L}$, we have $V_{0,y}^{R}=mV_{0,x}^{R}$ and $V_{0,y}^{L}=lV_{0,x}^{L}$. Moreover, the normal hopping $t^{\prime}_{0,x/y}$ can read as
\begin{equation}
 t^{\prime}_{0,x/y}= E_{R}^{x/y}\frac{4}{\sqrt{\pi}}\left(\widetilde{V}_{0,x/y}^{L}\right)^{3/4}\exp{\left[-2\sqrt{\widetilde{V}_{0,x/y}^{L}}\right]}
\end{equation}
which is the solution of the $1D$ Mathieu equation \cite{Tipical_Temperature}.

When $t_{y}=t_{x}$, the value of  $\delta_{t}/t_{x}$ is an important parameter which impacts the value of topological current $J_{\rm{D}}$. Thus, we show  $t_{y}=t_{x}$ and $\delta_{t}/t_{x}$ as a function of $l$ and $m$ for given $\widetilde{V}_{0,x}^{L}$ in  Figs.~\ref{tx_delta_app}. From Fig.~\ref{tx_delta_app}(a), the curve $t_{y}=t_{x}$ changing with the value of $\widetilde{V}_{0,x}^{L}$ is not distinct. So, we chose $\widetilde{V}_{0,x}^{L}=5$ as an example to show $\delta_{t}/t_{x}$ as a function of $l$ and $m$.  From  Figs.~\ref{tx_delta_app}, we easily know that $\delta_{t}/t_{x}=0.32$ with $t_{x}=t_{y}$ can be realized in the experiment. In order to clearly detect this novel topological current $J_{\rm{D}}$ under current experimental conditions, we want to observe the maximum drifting of the center mass $x_{\rm{c}}^{\rm{max}}$ within the coherent time $\tau_{\rm{coh}}$. This means that the measuring time $\tau\sim31.4 T_{0}$, which is the time for the center of mass moving from $|x_{\rm{c}}|^{\rm{min}}$ to $|x_{\rm{c}}|^{\rm{max}}$, must be smaller than $\tau_{\rm{coh}}\sim 100 \rm{ms}$. In Fig.~\ref{tx_ty_app}, we show the range of possible values of trapping potential $\widetilde{V}_{0,x}^{L}$ and Raman potential $\widetilde{V}_{0,x}^{R}$ for $^{87}\rm{Rb}$ [see Fig.~\ref{tx_ty_app}(a)] and $^{23}\rm{Na}$ [see Fig.~\ref{tx_ty_app}(b)]atom systems, in which the measuring time $\tau$ is smaller than the coherent time  $\tau_{\rm{coh}}$ and the drifting of the centre of mass $\Delta x_{c}$ is the largest. Furthermore, if $\rho <0.02 (a_{x}a_{y})^{-1}$,  $\Delta x_{c}$ can be detected for both cases (having one or two pairs of Dirac points), where the the measurement accuracy of $\Delta x_{c}$ is about $a_{x}$. For the case of $\rho=0.01 (a_{x}a_{y})^{-1}$, if the parameters of the systems are chosen with the same value of the parameters as the blue dashed line presented in Fig.~4, the experimenters can verify the $J_{D}$ existence within a very short time, $\tau_{\rm{min}}=2T_{0}$ (within $2T_{0}$,  the corresponding $\Delta x_{c}=a_{x}$). The range of possible values of the Raman potential $\widetilde{V}_{0,x}^{R}$ ($\widetilde{V}_{0,x}^{R}>0.18$ for $^{87}\rm{Rb}$  and $\widetilde{V}_{0,x}^{R}>0.05$ for $^{23}\rm{Na}$) for $^{87}\rm{Rb}$  and $^{23}\rm{Na}$ atom systems is dramatically expanded, in which the experimenters can verify the $J_{D}$ existence with the corresponding measuring time $\tau_{\rm{min}}\leq\tau_{\rm{coh}}$.

\begin{figure}[h!]
\includegraphics[width=0.9\linewidth]{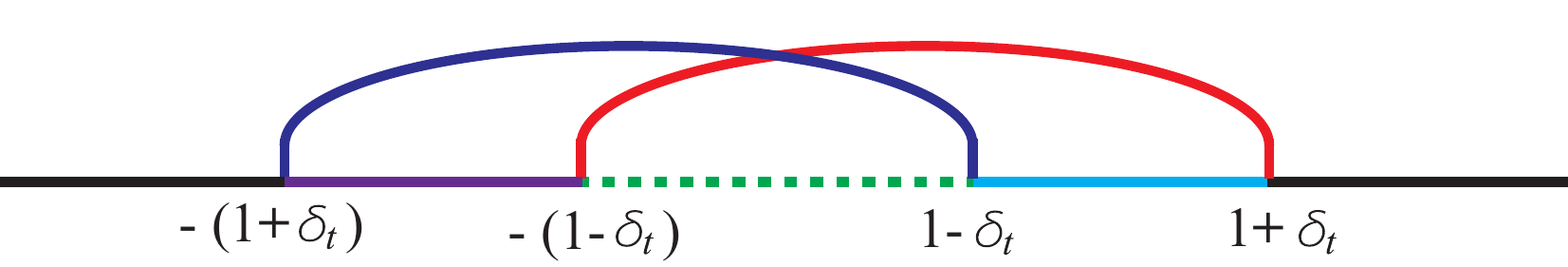}
\caption{The sketch of the solution of Eqs.~(\ref{equa_dirac_1}) and (\ref{equa_dirac_2})}
\label{Dirac_number}
\end{figure}

\section{Tuning the number of pairs of Dirac points} \label{App_Dirac}

For momentum-space Hamiltonian
\begin{eqnarray}
\mathcal{H}\left(k\right)&=&\sin\left(k_{x}a_{x}\right)\hat{\sigma}_{x}+\left[\lambda(\tau)-\delta_{t}\cos\left(k_{x}a_{x}\right)\right. \nonumber \\
&&-\cos\left(k_{y}a_{y}\right)\left.\right]\hat{\sigma}_{y} +f\left(\mathbf{k}\right)\hat{\sigma}_{0},\label{HK}
\end{eqnarray}
the Dirac points can be obtained by the solution of equations $\sin\left(k_{x}a_{x}\right) =0$ and
$\lambda(\tau)-\delta_{t}\cos\left(k_{x}a_{x}\right) -\cos\left(k_{y}a_{y}\right)=0$. It is easy to know the solutions in the $k_{x}$ direction, which can be written as $k^{D}_{x}=0,\pi$, and the $k_{y}$  direction solutions $k^{D}_{y}$ must satisfy the equations
\begin{equation}
 -1\leq \lambda  \left(\tau\right) -\delta_{t} \leq 1, \label{equa_dirac_1}
\end{equation}
and
\begin{equation}
 -1\leq  \lambda \left(\tau\right)+\delta_{t} \leq 1, \label{equa_dirac_2}
\end{equation}
with  $1>\delta_{t}>0$. It is found that the solution of Eq.~(\ref{equa_dirac_1}) is
\begin{equation}
 -\left(1-\delta_{t}\right)\leq \lambda \left(\tau\right) \leq 1+\delta_{t},  \label{equa_dirac_solution_1}
\end{equation}
and the solution of Eq.~(\ref{equa_dirac_2}) is
\begin{equation}
 -\left(1+\delta_{t}\right)\leq  \lambda \left(\tau\right) \leq 1-\delta_{t}. \label{equa_dirac_solution_2}
\end{equation}

There are four different cases, which are presented at follows: in the first case $\lambda \left(\tau\right)>1+\delta_{t}$ or $\lambda \left(\tau\right)<-(1+\delta_{t})$ (see the region of the black lines in Fig.~\ref{Dirac_number}), there are no Dirac points in systems. In the second case $ (1-\delta_{t})<\lambda\left(\tau\right)\leq (1+\delta_{t})$ (see the region of the cyan line in Fig.~\ref{Dirac_number}), there is only one pair of Dirac points, which is denoted as $k_{D,1}=\left(0,\pm k_{y,1}\right)$. In the third case $ -(1-\delta_{t})<\lambda \left(\tau\right)<(1-\delta_{t})$ (see the region of the green dashed line in Fig.~\ref{Dirac_number}), there are two pairs of  Dirac points, which are denoted as $k_{D,1}^{\pm}=\left(0,\pm k_{y,1}\right)$ and $k_{D,2}^{\pm}=\left(\pi,\mp k_{y,2}\right)$. In the fourth case $ -(1+\delta_{t})<\lambda\left(\tau\right)<-(1-\delta_{t})$ (see the region of the purple line in Fig.~\ref{Dirac_number}), there is  only one pair of Dirac points, which is denoted as $k_{D,2}=\left(\pi,\pm k_{y,2}\right)$. Thus, the number of  pairs of Dirac points can be regulated by changing the value of $\lambda\left(\tau\right)$. In Figs.~\ref{Raman-laser}(b) and ~\ref{Raman-laser}(c), we choose two typical points, $\delta_{t}=0.32$, and $\lambda\left(\tau\right)=1.2$ or $\lambda\left(\tau\right)=0.48$, as two examples to reveal this characteristic.

\end{appendix}

\end{document}